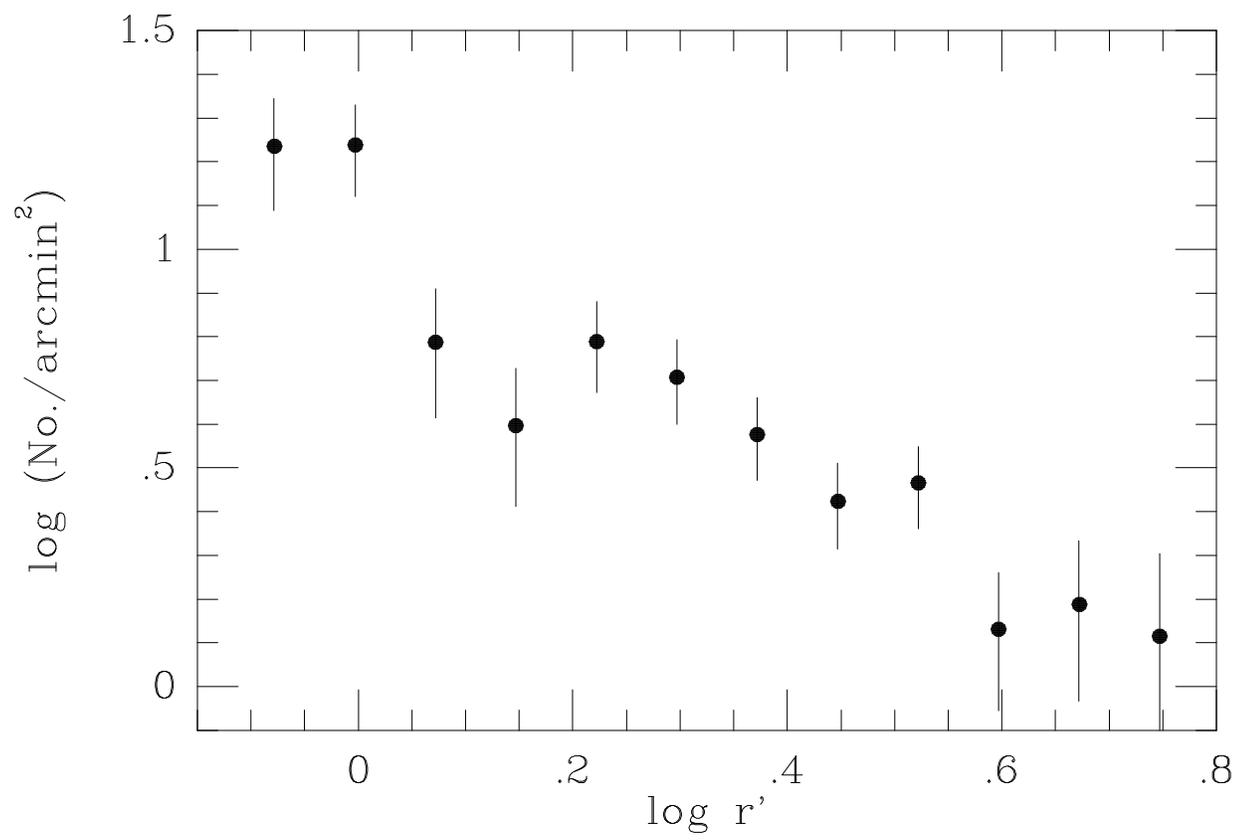

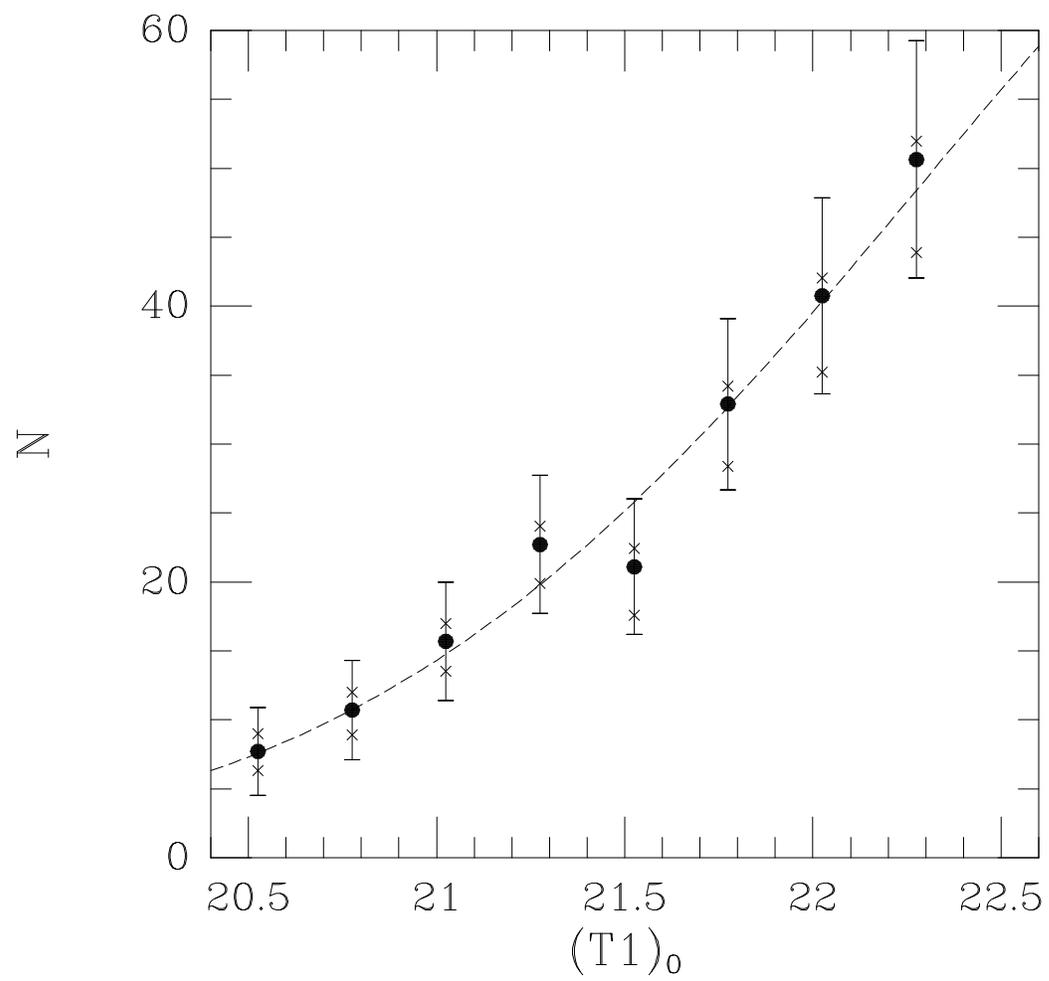

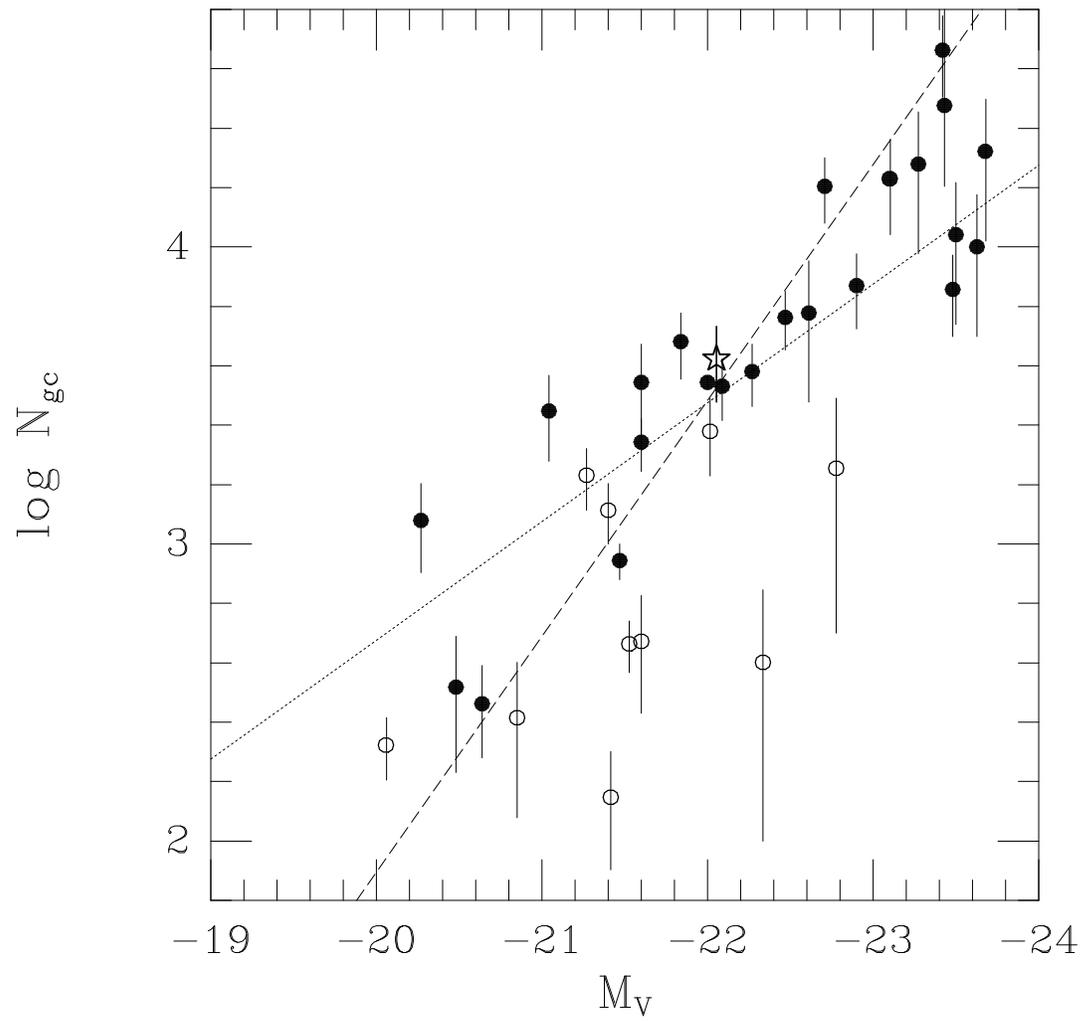

# THE RICHNESS OF THE GLOBULAR CLUSTER SYSTEM OF NGC 3923: CLUES TO ELLIPTICAL GALAXY FORMATION

Stephen E. Zepf [1]

*Department of Astronomy, University of California*

*Berkeley, CA 94720*

*e-mail: zepf@astron.berkeley.edu*

Doug Geisler

*Cerro Tololo Inter-American Observatory*

*National Optical Astronomy Observatories* [2]

*Casilla 603, La Serena, Chile*

*e-mail: dgeisler@ctio.noao.edu*

Keith M. Ashman

*Department of Physics and Astronomy, University of Kansas*

*Lawrence, KS 66045*

*ashman@kusmos.phsx.ukans.edu*

**Abstract**

We present new data on the globular cluster system of the elliptical galaxy NGC 3923 which show that it has the most globular clusters per unit luminosity of any non-cluster elliptical yet observed, with $S_N = 6.4 \pm 1.4$. NGC 3923 is also among the brightest ellipticals outside of a galaxy cluster for which the number of globular clusters has been determined. Our observation of a large number of clusters per unit luminosity (high $S_N$ value) for a bright elliptical in a sparse environment is consistent with the suggestion of Djorgovski & Santiago that the number of globular clusters is a power law function of the luminosity with an exponent greater than one. We relate this higher specific frequency of globular clusters in more luminous galaxies to other observations which indicate that the physical conditions within elliptical galaxies at the time of their formation were dependent on galaxy mass.

Subject headings: galaxies: star clusters - galaxies: formation - galaxies: elliptical and lenticular

## 1. Introduction

The presence of globular clusters (GCs) around nearly all galaxies suggests that the formation of globular clusters is closely tied to the formation of galaxies themselves. Furthermore, the properties of GC systems are clearly correlated with the properties of their parent galaxies. As reviewed by Harris (1991), one of the first correlations to be recognized was the tendency for brighter galaxies to have more globular clusters. This led to the introduction of the specific frequency, $S_N$, which is defined as the number of globular clusters ($N_{gc}$) per unit ($M_V = -15$) galaxy luminosity (Harris & van den Bergh 1981). The number of GCs is also correlated with galaxy morphology, with elliptical galaxies having more GCs per unit luminosity (higher $S_N$ values) than spiral galaxies.

Variations in the $S_N$ values of elliptical galaxies are significantly greater than the estimated measurement errors. These differences are observed among ordinary elliptical galaxies and are not dependent on the well-known high $S_N$ values of some cD galaxies. Harris (1991) and others have suggested that at least some of this variation can be accounted for by a dependence of $S_N$ on environment, with ellipticals in clusters tending to have more GCs per unit luminosity than ellipticals in small groups. Starting with the same data, Djorgovski & Santiago (1992) proposed the primary correlation is with luminosity. Specifically, they found that $N_{gc}$ scales with luminosity roughly as $N_{gc} \propto L^2$, and thus $S_N$ increases with galaxy luminosity.

One of the reasons that two very different hypotheses have been proposed on the basis of the same data is that most of the bright ellipticals for which $N_{gc}$ has been determined are in galaxy clusters. Therefore, we undertook a study of the GC system of NGC 3923, which is a bright elliptical in a small group. The absolute magnitude of NGC 3923 is $M_V = -22.05$, based on $B_T^0 = 10.52$, $(B-V)_0 = 0.95$, and the $D_n$-$\sigma$ distance of 1587 km s$^{-1}$ from Faber et al. (1989), and $H_0 = 75$ km s$^{-1}$ Mpc$^{-1}$, which we adopt throughout this letter. NGC 3923 is located in a low-density region, and is best-known for its shells, a common feature of ellipticals outside of galaxy clusters (Schweizer & Seitzer 1992). In all other respects, it appears to be an ordinary elliptical galaxy.

In this letter, we determine the number of GCs and $S_N$ value of NGC 3923, and discuss the implications of this observation for the formation of both elliptical galaxies and their globular clusters. The observations and data reduction are described in Section 2, followed in Section 3 by the determination of $N_{gc}$ and $S_N$ for NGC 3923. In Section 4 we compare the results for NGC 3923 to those for other ellipticals and link these data to other observational constraints on the formation of elliptical galaxies.

## 2. Observations and Data Reduction

The data from which we determine the number of globular clusters associated with NGC 3923 are deep images obtained at the prime focus of the 4m at CTIO, using the Tek1024 CCD on UT 20 Feb 1993. We utilized the $C$ and $T_1$ passbands of the Washington photometric system, obtaining $7 \times 1000$s images in $C$ and $5 \times 1000$s in $T_1$. This system is useful for our purpose since it combines wide bandpasses necessary for observing faint objects with a broad wavelength baseline for estimating metallicities (Geisler & Forte 1990). The images were processed in the standard way within IRAF, and then registered and median combined. Stellar images in the combined frames have a FWHM of $\sim 1.5''$. The photometry of the objects in the field was performed with the version of DAOPHOT in IRAF, using an iterative process to remove the background light of the galaxy (e.g. Fischer et al. 1990).

We examined the list of objects photometered by DAOPHOT in order to distinguish background galaxies and foreground stars from the GCs of NGC 3923. Since GCs are point sources at the distance of NGC 3923, background galaxies can be removed from the sample based on their extended nature. We employed a combination of the image shape estimators in the DAOPHOT package and those described by Harris et al. (1991) to separate background galaxies from our GC sample. Image classification cannot distinguish stars from the GCs of NGC 3923. However, the majority of the foreground stars are dwarfs in the Galactic disk, which are much redder than any GC and can be removed by the use of a color criterion. By limiting the colors of objects in our final sample to $1.05 < (C - T_1)_0 < 2.15$, we eliminate most foreground stars but only a minimal fraction of GCs. In a study similar to ours, Couture et al. (1991) have argued that the combination of image classification and color cuts results in a sample composed of $> 90\%$ GCs.

In order to determine the completeness of our sample as a function of $C$ and $T_1$ magnitude, we performed extensive artifical star tests. We find that for a typical GC color of $(C - T_1)_0 = 1.63$ (E(B-V) = 0.06, Burstein & Heiles 1982) the completeness is 90% at $(T_1)_0 = 21.98$, and 50% at $(T_1)_0 = 22.51$. However, our combined $C$ image is shallower than our $T_1$ image, so the incompleteness at a given $T_1$ magnitude is larger for red objects than for blue ones. For example, at the red limit of the GC sample, $(C - T_1)_0 = 2.15$, the completeness is 90% at $(T_1)_0 = 21.58$, and 50% at $(T_1)_0 = 22.10$. For our final sample, we select the magnitude limits $20.4 < (T_1)_0 < 22.4$. The bright limit is set by the onset of the NGC 3923 GC population, and the faint limit insures $> 50\%$ completeness over essentially the entire sample, while maximizing the sample size. The typical incompleteness in this sample is about 10%.

The final sample is thus composed of the 190 point sources with $20.4 < (T_1)_0 < 22.4$, $1.05 < (C - T_1)_0 < 2.15$ and which are within the elliptical area $55.2'' < \sqrt{(ab)} < 230.0''$, where the ellipticity is 0.4 and $a$ and $b$ are the semi-major and semi-minor axes respectively, and the total usable area is 35.0 arcmin$^2$. We plot the radial profile of this sample in Figure 1. The radial profile and ellipticity are discussed more fully in Zepf, Geisler, & Ashman (1994). As noted above, the contamination in this sample is expected to be small. The contribution from foreground stars predicted by the Bahcall & Soneira model of the Galaxy (Bahcall 1986) is 10.4, or 5% of the total sample. As an upper limit, we adopt a contamination level of 20%, which accounts for the highest contribution estimated from studies of galaxy counts. This would require almost half of the background galaxies to be misclassified as stellar, which is many more than expected from previous studies of image classification (e.g. Harris et al. 1991). For a lower limit, we simply assume no contamination. We adopt 5% as our best estimate of the contamination level, as it is the most consistent with other studies of this type, although we include calculations using all three levels (0%, 5%, and 20%) to show how uncertainty in the background level affects the determination of $N_{gc}$. A more detailed account of this analysis is given in Zepf et al. (1994).

### 3. The Number of Globular Clusters Associated with NGC 3923

The standard method for determining the total number of globular clusters associated with a galaxy is based on fitting the observed number of GCs as a function of magnitude, N($m$), to an assumed form of the globular cluster luminosity function (GCLF). For the latter, we adopt the standard, N($m$) = $N_0 e^{-(m-m_0)^2/2\sigma^2}$, where $N_0$ is a scale factor, $m_0$ is the peak or "turnover" magnitude of the GCLF and $\sigma$ is the dispersion. Since our data only reach to about 1.5 mag brighter than the expected peak in the GCLF, we cannot reliably make independent determinations of the three parameters, $N_0$, $m_0$, and $\sigma$ (e.g. Hanes & Whittaker 1987). Fortunately, there are other constraints on both $\sigma$ and $m_0$. For $m_0$, a value of $23.8 \pm 0.5$ mag is found by combining the distance modulus of $31.6 \pm 0.4$, based on the $D_n - \sigma$ distance of NGC 3923 given by Faber et al. (1989), and the GCLF peak of $-7.8 \pm 0.2$ mag in $T_1$, derived by transforming the V peak given by Harris (1991) to $T_1$. Harris et al. (1991) found $\sigma = 1.46 \pm 0.07$ for the combined GCLF of three ellipticals in Virgo, which represents the best database to date for galaxies similar to NGC 3923. A somewhat wider range of values has been found for other types of galaxies, including $\sigma = 1.2$ for the Galaxy and $\sigma = 1.7$ for M87 (e.g. McLaughlin et al. 1994).

We explore this parameter space by considering three values for $m_0$: 23.3, 23.8 and 24.3. These bracket the likely range of this parameter. We fix $m_0$ at each of these values and fit the GCLF to find $N_0$ and $\sigma$. The resulting values of $\sigma$ are $1.33 \pm 0.08$, $1.51 \pm 0.10$ and $1.66 \pm 0.09$, respectively. Each of the $m_0$, $\sigma$ combinations fits the data well. However, it is notable that the fit with the preferred value of $m_0 = 23.8$ also gives a value of $\sigma$ closest to that typically found for normal elliptical galaxies. The GCLF of NGC 3923 and the fit to the data for $m_0 = 23.8$ are plotted in Figure 2. The error bars have been computed by convolving the Poisson uncertainties in the number counts with the incompleteness function (Bolte 1989). Also shown as crosses above and below each point are values corresponding to the lower and upper limits to the background.

The final step necessary to determine the total number of GCs is the extrapolation from the radial region covered by our data to the entire GC system of NGC 3923. We first fit the radial falloff of the surface density of GCs to both a power law in $r$ and a deVaucouleurs $r^{1/4}$ profile. Each form provides a fairly good fit to the data, with a $\chi^2$ per degree of freedom of 1.3. We then determine the total number of GCs by integrating these profiles from the center of the galaxy out to large radii. For the inner radial limit, we use 2 kpc, which is roughly the limit inside of which dynamical forces are expected to be efficient at destroying GCs (Harris 1988 and references therein). A factor of two variation in this inner cut only changes the total number of GCs by several percent. An outer cutoff is not required for the $r^{1/4}$ profile, which converges at large radii. However, the power law fit diverges if integrated to infinity, so we adopt an outer limit of 100 kpc. The resulting total numbers for the two profile shapes differ by only a few percent. These calculations show that about $\sim 30\%$ of the total population is located within our radial region.

We can now combine these factors to derive $N_{gc}$ and $S_N$ for NGC 3923. In Table 1, we present the results for the various combinations of $m_0$, $\sigma$, and background level. The preferred combination of $(m_0, \sigma) = (23.8, 1.51)$ and the intermediate background level gives $N_{gc} \simeq 4300$, and the range of $N_{gc}$ from acceptable fits to the GCLF is roughly 2000 to 7000. The value of $S_N$ ranges from 4.4 to 7.3, with the preferred parameters giving

$S_N = 6.4$.

## 4. Discussion

We compare the richness of the NGC 3923 GC system to that of the GC systems of other elliptical and cD galaxies in Figure 3, which is a plot of $M_V$ vs. $\log N_{gc}$ for NGC 3923 and the elliptical and cD galaxies in the compilation of $N_{gc}$ values of Harris (1994). $M_V$ is determined from the photometry and distances of Faber et al. (1989) when available and from the RC3 for the few very distant cDs for which it is not, except for the magnitude of NGC 3311, which is taken from Hamabe (1993). This plot shows that NGC 3923, a bright elliptical in a sparse environment, has a GC system which is richer than any of the generally less luminous, non-cluster ellipticals previously observed, and is similar in richness to the GC systems of cluster ellipticals of comparable luminosity. The NGC 3923 GC system falls along the general trend of increasing $N_{gc}$ with increasing $L$. Specifically, the location of NGC 3923 in Figure 3 is consistent with the suggestion of Djorgovski & Santiago (1992) that the number of GCs scales with host galaxy luminosity to a power greater than one.

The scaling of $N_{gc}$ with $L$ to a power greater than one implies that the efficiency of GC formation relative to currently luminous stars increases with host galaxy luminosity. When combined with a consideration of the physical conditions favorable for GC formation, the relationship between $N_{gc}$ and $L$ provides important constraints on formation of elliptical galaxies. Although the understanding of GC formation is incomplete, observations have revealed newly formed GCs in several galaxies which are undergoing tidal interactions or mergers (Holtzman et al. 1992, Whitmore et al. 1993, Conti & Vacca 1994). These discoveries appear to confirm earlier predictions that mergers of gas-rich galaxies are favorable sites for GC formation (Schweizer 1987, Ashman & Zepf 1992). In particular, the physical conditions in these systems may promote the build up of very massive, dense molecular clouds, which are likely to be the progenitors of GCs (Larson 1990, Ashman & Zepf 1992,

Kumai, Basu, & Fujimoto 1993, Harris & Pudritz 1994). Therefore, the implication of our study of GC systems is that higher luminosity ellipticals formed in conditions more similar to those in mergers of gas-rich galaxies.

The constraints on the conditions during the formation of elliptical galaxies are much stronger when GC studies are combined with other observations of elliptical galaxies. Specifically, the physical conditions suggested by our GC results are intriguingly similar to those suggested by the observation that the abundance ratio of Mg to Fe becomes progressively larger for more luminous and massive ellipticals (Worthey, Faber, & González 1992, Peletier 1989). As discussed by Worthey et al., the differences in abundance ratios as a function of elliptical galaxy mass indicate either - 1) the star formation epoch of massive ellipticals was short compared to the timescale of type I supernovae, 2) the IMF in brighter ellipticals was enhanced in massive stars, producing additional type II supernovae and thus more Mg, or 3) Fe is selectively lost in the galactic winds of more massive ellipticals. The third possibility seems unlikely since it is difficult to arrange for the required mass-dependent selective loss of metals. Extremely rapid star formation in elliptical galaxies as a whole is inconsistent with the observed shape of the metallicity distribution of the GC systems, which favors episodic formation in two or more bursts rather than a single, rapid formation event (Zepf & Ashman 1993, Zepf, Ashman, & Geisler 1994).

These results point to a picture in which individual gas-rich galaxies merge to form the initial giant elliptical, with the mergers being more gas rich or more efficient at producing high gas densities in the progenitors of brighter and more massive ellipticals. Both theoretical arguments (e.g. Silk 1993), and observations of nearby galaxy starbursts and mergers (e.g. Joseph 1991, Rieke 1991) suggest that these physical conditions may lead to a stellar IMF which is enhanced in high mass stars. Thus the merger of gas-rich systems is able to account for the trend of both the number of GCs and the abundance ratios as a function of elliptical galaxy mass. The relationship $(M/L) \propto M^{1/6}$, which follows from the Fundamental Plane of elliptical galaxies (Faber et al. 1987), might also be at least partially explained by a population of remnants of high mass stars which are more common in massive ellipticals. The decrease in central density with increasing elliptical galaxy mass (Kormendy 1989) may result from subsequent mergers of primarily stellar systems, which formed by previous gas-rich mergers, as suggested by Bender, Burstein, & Faber (1992).

We thank the referee, Bill Harris, for many useful suggestions. S.E.Z. acknowledges support from NASA through grant number HF-1055.01-93A awarded by the Space Telescope Science Institute, which is operated by the Association of Universities for Research in Astronomy, Inc., for NASA under contract NAS5-26555. K.M.A. acknowledges support from a Fullam/Dudley Award and a Dunham Grant from the Fund for Astrophysical Research.

TABLE 1: Fits to the GCLF of NGC 3923

| $m_0$ | $\sigma$ | bckg | $N_{gc}$ | $S_N$ |
|---|---|---|---|---|
| 23.8 | 1.51 | standard | 4261 | 6.4 |
| 23.8 | 1.51 | high | 2992 | 4.5 |
| 23.8 | 1.51 | low | 4845 | 7.3 |
| 23.3 | 1.33 | standard | 2627 | 6.3 |
| 23.3 | 1.33 | high | 1845 | 4.4 |
| 23.3 | 1.33 | low | 2986 | 7.2 |
| 24.3 | 1.66 | standard | 6623 | 6.3 |
| 24.3 | 1.66 | high | 4646 | 4.4 |
| 24.3 | 1.66 | low | 7524 | 7.2 |

# Figure Captions

Figure 1 - The radial profile of the GC system of NGC 3923. In order to better describe the radial distribution of GCs, the profile shown above extends to larger radii than the sample used in this work. The error bars are the Poissonian uncertainties of the number of objects detected in each bin.

Figure 2 - A plot of the number of GCs as a function of magnitude for NGC 3923. The filled circles represent the number using the standard background value, with the crosses representing the numbers using the upper and lower limits for the background. The dashed line is the fit to these data with $m_0 = 23.8$, $\sigma = 1.51$ and the standard Gaussian GCLF. The other $m_0$, $\sigma$ combinations discussed in the text are indistinguishable from this fit within the magnitude range of our data.

Figure 3 - The log of the total number of globular clusters plotted against the absolute magnitude of the parent galaxy. The solid dots are galaxies in clusters, the open circles are non-cluster galaxies, and the star symbol represents NGC 3923. The dashed line indicates the slope expected for $N_{gc} \propto L^2$, and the dotted line corresponds to $N_{gc} \propto L^1$. A weighted least squares fit to these data lies between these two cases, although no single power law provides a good fit, perhaps reflecting flattening in the trend at very high luminosities.